\documentclass[12pt,preprint]{aastex}

\newcommand{\ba}{\begin{eqnarray}}
\newcommand{\ea}{\end{eqnarray}}
\def\ra{\rightarrow}
\def\eps{\epsilon}
\def\vareps{\varepsilon}

\begin{document}

\title{MeV-GeV emission from neutron-loaded short gamma-ray burst jets}

\author{Soebur Razzaque and Peter M\'esz\'aros}
\affil{Department of Astronomy \& Astrophysics, Department of Physics, 
Pennsylvania State University, University Park, PA 16802 \\
soeb@astro.psu.edu, nnp@astro.psu.edu}

\begin{abstract}
Recent discovery of the afterglow emission from short gamma-ray bursts
suggests that binary neutron star or black hole-neutron star binary
mergers are the likely progenitors of these short bursts. The
accretion of neutron star material and its subsequent ejection by the
central engine implies a neutron-rich outflow. We consider here a
neutron-rich relativistic jet model of short bursts, and investigate
the high energy neutrino and photon emission as neutrons and protons
decouple from each other. We find that upcoming neutrino telescopes
are unlikley to detect the 50 GeV neutrinos expected in this
model. For bursts at $z\sim 0.1$, we find that {\it GLAST} and
ground-based Cherenkov telescopes should be able to detect prompt 100
MeV and 100 GeV photon signatures, respectively, which may help test
the neutron star merger progenitor identification.
\end{abstract}


\section{Introduction}

The nature and origin of short gamma-ray bursts (SGRBs), with
$t_\gamma \lesssim 2$ s duration, were largely conjectural until
recently. This changed dramatically with the detection of the
afterglow and the identification of the host galaxy of GRB 050509b at
a redshift $z=0.226$ \citep{getal05}, as well as subsequent
observations of other short burst afterglows, e.g., GRB 050709
\citep{fetal05, vetal05}, GRB 050724 \citep{betal05}, etc. These
observations revealed that the short bursts are located mainly in
elliptical hosts, while some are in spirals or irregulars, as expected
from an old population such as NS-NS or NS-BH mergers. They also
provided, for the first time, details about the X-ray, and in some
cases optical and radio afterglow lightcurves, including in two cases
evidence for a jet break, qualitatively similar to the relativistic
jet afterglows of long GRBs [see, e.g., \cite{zm04} for reviews of
long GRBs]. The average redshifts are lower than for long bursts,
indicating an isotropic-equivalent energy of SGRBs approximately two
orders of magnitude lower than for long ones.

We investigate here a model of SGRB jet which is initially loaded with
neutrons and fewer protons, as expected from a neutron star
composition.  High energy emission in this model is then expected from
the photosphere of the relativistic jet outflow, including both
thermal and non-thermal components, in addition to the radiation
expected outside of it.  We describe our jet model in Section 2, high
energy emission processes in Section 3, and expected neutrino and
photon fluxes in Section 4 and 5 respectively. We summarize and
discuss our results in Section 6.

\section{Neutron loaded jet dynamics}

We assume the total isotropic-equivalent energy outflow in the SGRB
jet of $L=10^{50} L_{50}$ erg/s. The jet is loaded with protons and
neutrons with an inital neutron to proton number density ratio $\xi_o
=n'_n/n'_p = 10 \xi_{o,1}$ in the comoving frame.\footnote{We use
primed variables in the relativistic plasma frame and un-primed
variables in the observer frame or laboratory frame} The total mass
outflow rate in such a neutron-rich jet is ${\dot M} = 4\pi r^2 c
(1+\xi_o) n'_p m_p$ initially, since $m_p \simeq m_n \gg m_e$ and the
thermal energy is negligible. A useful quantity is the usual total
energy to mass flow ratio $\eta=L/{\dot M}c^2$, characterizing the
dimensionless entropy or the baryon loading of the jet. In a typical
long GRB ($t_\gamma \gtrsim 2$ s) the final bulk Lorentz factor of the
$\gamma$-ray emitting fireball is equivalent to $\eta$. In the case of
SGRBs, we assume $\eta = 316\eta_{2.5}$, similar to that of the long
GRBs, motivated by observation.

The outflow starts at a radius $R_o = 10^6 R_{o,6}$ cm which is a few
times the Schwarzschild radius $r_g = 2GM_{\rm bh}/c^2$ of a solar
mass black hole created by binary mergers. The inital temperature of
the plasma outflow is
\ba T'_o = \left( \frac{L}{4\pi R_o^2 c a} \right)^{1/4} \approx
1.2 \left( \frac{L_{50}}{R_{o, 6}^2}\right)^{1/4} ~{\rm
MeV},
\label{initial-plasma-temp} \ea
where $a= \pi^2 k^4/15(\hbar c)^3 = 7.6\times 10^{-15}~{\rm erg}~{\rm
cm}^{-3}~{\rm K}^{-4}$ is the radiation density constant. The
optically thick hot plasma expands adiabatically due to the radiation
pressure. The comoving temperature drops as $T'(r) \propto R_o/r$ and
the plasma expands with an increasing bulk Lorentz factor $\Gamma(r)
\propto r/R_o$ with the radius $r$ following the adiabatic law. 
Electrons (both $e^\pm$ and $n_e \simeq n_p$) are coupled to the
photons via Compton scattering, and protons, which are coupled to the
electrons via Coulomb interaction, are held together in the expanding
optically thick plasma. Neutrons, however, are somewhat loosely
coupled to the protons by elastic nuclear scattering with a
cross-section $\sigma_{np} \approx 3\times 10^{-26}$ cm$^2$, which is
roughly a fraction 1/20 of the Thomson cross-section ($\sigma_{\rm Th}
\approx 6.65\times 10^{-25}$ cm$^2$). This changes considerably the
dynamics of a neutron-rich jet [e.g. \cite{dkk99, bm00, b03}] as we
discuss next.

The protons and neutrons are coupled together with a common bulk
Lorentz factor $\Gamma_n (r)\simeq \Gamma_p (r)\simeq \Gamma (r)$ as
long as the $n$-$p$ collision time $t'_{np} \simeq (n'_p
\sigma_{np}c)^{-1}$ is shorter than the plasma expansion time $t'_{\rm
exp} \simeq r/c\Gamma (r)$. The critical dimensionless entropy for
which $n$-$p$ decoupling happens, from the condition $t'_{np} =
t'_{\rm exp}$, is
\ba \eta_{np} \simeq \left[ \frac{L\sigma_{np}}{4\pi R_o m_p c^3 
(1+\xi_o)} \right]^{1/4} \approx 150 \left[ \frac{L_{50}}{R_{o,6}
(1+\xi_{o,1})} \right]^{1/4}. \label{np-entropy} \ea
As a result, the neutrons and protons in our SGRB jet model ($\eta
\gtrsim \eta_{np}$) decouple at a radius where the nuclear scattering 
optical depth $\tau'_{np} \simeq n'_p \sigma_{np}
R_{np}/\Gamma(R_{np}) =1$ as
\ba R_{np} \simeq \left[ \frac{L\sigma_{np}}{4\pi R_o 
\eta m_p c^3 (1+\xi_o)} \right]^{1/3} R_o = 
\eta_{np} \left(\frac{\eta_{np}}{\eta} \right)^{1/3}
\approx 1.2\times 10^8 \left[ \frac{L_{50} R_{o,6}^2} 
{\eta_{2.5} (1+\xi_{o,1})} \right]^{1/3} ~{\rm cm}.
\label{np-radius} \ea
The bulk Lorentz factor of the neutrons does not increase further and
saturates to a final value of
\ba \Gamma_{n,f} = \frac{R_{np}}{R_o} \approx 115\left[ \frac{L_{50}} 
{\eta_{2.5} R_{o,6} (1+\xi_{o,1})} \right]^{1/3},
\label{final-neutron-bulk} \ea
at $r=R_{np}$. The neutrons start to lag behind the protons at $r
\gtrsim R_{np}$, and the total energy outflow due to protons is
\ba {\hat L} = L- \Gamma_{n,f} {\hat {\dot M}}c^2 \xi_o = 
L \left(1- \frac{\Gamma_{n,f}}{\eta} \frac{\xi_o}{1+\xi_o}\right) 
\approx 7\times 10^{49} ~{\rm ergs/s}, \label{proton-flow} \ea
which is close to the initial value of $L$ since $\Gamma_{n,f} <
\eta$. Here ${\hat {\dot M}} \simeq {\dot M}/(1+\xi_o)$ is the proton
mass outflow rate after decoupling. The outflow is still optically
thick at $r=R_{np}$ since $\sigma_{\rm Th} \simeq 20 \sigma_{np}$. The
decoupling of radiation from plasma electrons takes place for a
critical value for the dimensionless entropy, for which the Thomson
scattering time scale $t'_{\rm Th} \simeq (n'_e \sigma_{\rm Th}
c)^{-1}$ is equal to the plasma expansion time, is given by
\ba {\hat \eta}_{\rm rad} =  \left( \frac{{\hat L} \sigma_{\rm Th}}
{4\pi R_o m_p c^3} \right)^{1/4} \approx 530 \left( 
\frac{{\hat L}_{49.8}} {R_{o,6}} \right)^{1/4}, 
\label{rad-entropy} \ea
for our SGRB parameter ${\hat L}_{49.8} = {\hat L}/(7\times 10^{49} ~
{\rm ergs/s})$. Note that the dimensionless entropy of the proton
outflow ${\hat \eta} \equiv {\hat L}/{\hat {\dot M}}c^2$ is related to
${\hat \eta}_{\rm rad}$ by
\ba \frac{{\hat \eta}}{{\hat \eta}_{\rm rad}} \simeq 
\left[ 1+\xi_o \left(1 - 
\frac{\Gamma_{n,f}}{\eta} \right) \right]^{3/4} 
\left( \frac{\sigma_{np}}{\sigma_{\rm Th}} \right)^{1/4} 
\frac{\eta}{\eta_{np}} \approx 4.4. \label{entropy-ratio} \ea
In this case (${\hat \eta} > {\hat \eta}_{\rm rad}$) the outflow
becomes transparent during the acceleration phase $\Gamma\propto r$,
at a radius $r=R_{\tau}$ where the Thomson optical depth $\tau'_{\rm
Th} \simeq n'_e R_{\tau}/\Gamma (R_{\tau}) =1$,
\ba R_{\tau} \simeq R_o {\hat \eta}_{\rm rad} 
\left( \frac{{\hat \eta}_{\rm rad}}{\hat \eta} \right)^{1/3} 
\approx 3.2\times 10^8 ~{\rm cm}, \label{transparent-radius} \ea
which is close to $R_{np}$. The final Lorentz factor of the (proton)
outflow is given by \citep{rbr05}
\ba \Gamma_{p,f} \simeq {\hat \eta}_{\rm rad} 
\left(\frac{\hat \eta} {{\hat \eta}_{\rm rad}} \right)^{1/9} \approx 624
\label{final-proton-bulk} \ea
which is achieved at a radius $r\sim R_{\tau}$ for this $\eta$. This
suggests a possible way to generate the high bulk Lorentz factor
estimated in short GRBs without any spectral lags in the data
\citep{nb06}.

Note that for $\eta<\eta_{np}$, one has $\Gamma_{n,f} =\Gamma_{p,f}
=\eta$ for a baryon loaded jet outflow satisfying the condition $\eta
< \eta_{\rm rad}$. Here $\eta_{\rm rad}$ is the initial critical
radiation entropy for combined proton and neutron outflow, defined
with an equation similar to equation (\ref{rad-entropy}) as
\ba \eta_{\rm rad} =  \left[ \frac{L \sigma_{\rm Th}}
{4\pi R_o m_p c^3 (1+\xi_o)} \right]^{1/4} \approx 322 \left[
\frac{L_{50}}{R_{o,6} (1+\xi_{o,1})} \right]^{1/4}
\label{rad-entropy-def} \ea

We have plotted the final proton and neutron bulk Lorentz factors as
functions of the baryon-loading parameter in Figure 1. For a given
total energy outflow rate $L$, the maximum allowed value of $\eta$,
from equation (\ref{rad-entropy-def}), are shown as three vertical
solid lines for three values of initial neutron to proton ratio $\xi_o
=$1, 5 and 10. Note that for $\eta \sim 316 \eta_{2.5}$, a typical
value used for long GRB modelling, the neutron and proton components
always decouple [see equation (\ref{np-entropy})] in case of SGRBs
because of a lower $L$. On the contrary, $n$-$p$ decouple only for
$\xi_o \gtrsim 4$ in case of a long GRB with $\eta \sim 316
\eta_{2.5}$, $L=10^{52} L_{52}$ erg/s and $R_o = 10^7R_{o,7}$ cm. 
Also note the high $\Gamma_{p,f}$ values of 519, 572 and 624 for
$\xi_o =$ 1, 5 and 10 respectively in Figure 1 from equation
(\ref{final-proton-bulk}).

\section{Inelastic neutron-proton scattering}

The elastic component of the total $np$ cross-section, which dominates
at lower energy, drops rapidly above the pion production threshold of
$\approx 140$ MeV, where the inelastic $np$ scattering cross-section
is $\sigma_{\pi} \approx \sigma_{np}$. Hence, the condition $\eta
\gtrsim \eta_{np}$ (and subsequently ${\hat \eta} \gtrsim 
{\hat \eta}_{\rm rad}$) for $n$-$p$ decoupling implies that there will
be pion production by the threshold processes, with roughly equal
(total unit) probability to produce a $\pi^\pm$ pair or (and)
$2\pi^0$,
\ba p n \ra \cases{ 
nn\pi^+ ~;~ \pi^+ \ra \mu^+ \nu_{\mu} \ra e^+ \nu_e {\bar \nu}_{\mu}
\nu_{\mu} \cr pp\pi^- ~;~ \pi^- \ra \mu^- {\bar \nu}_{\mu} \ra e^- 
{\bar \nu}_e \nu_{\mu} {\bar \nu}_{\mu} \cr pn\pi^0 ~;~ \pi^0 \ra 
\gamma \gamma } \label{pi-processes} \ea
The secondary pions produced by the inelastic $np$ scattering are at
rest in the center-of-mass (c.m.) frame of interaction. However, the
neutrons are not cold in the comoving outflow frame as we discuss
next, and the c.m. frame acquires a velocity relative to this frame.

The Lorentz factor of the neutrons relative to the outflowing protons,
at the radius $r\sim R_{\tau} \sim R_{np}$, is given by
\ba \Gamma_{\rm rel} = \Gamma_p \Gamma_n (1-\beta_p\beta_n)
\simeq \frac{1}{2} \left( \frac{\Gamma_{p,f}}{\Gamma_{n,f}} 
+ \frac{\Gamma_{n,f}}{\Gamma_{p,f}} \right) \approx 2.8
\label{n-p-bulk} \ea
Thus, in (proton) outflow rest frame the neutron energy is $\eps'_n =
\Gamma_{\rm rel}m_n c^2$ and the Lorentz factor of the c.m. is given by
$\gamma'_{\rm c.m.} = (\eps'_n +m_p c^2)/ (m_p^2c^4 +m_n^2c^4 +2
\eps'_n m_pc^2)^{1/2} \approx 1.4$. The energies of the decay
neutrinos are $\eps'_{\nu_{\mu}} \simeq \eps'_{{\bar \nu}_{\mu}}
\approx 30\gamma'_{\rm c.m.}$ MeV from $\pi^\pm$, $\eps'_{\nu_e} \simeq
\eps'_{{\bar \nu}_e} \approx 30\gamma'_{\rm c.m.}$ MeV and
$\eps'_{\nu_{\mu}} \simeq \eps'_{{\bar \nu}_{\mu}}
\approx 50\gamma'_{\rm c.m.}$ MeV from $\mu^\pm$. The $\pi^0$ decay 
photon energy is $\eps'_{\gamma} \approx 70\gamma'_{\rm c.m.}$ MeV.
The observed energies of the decay products would be
\ba \eps = \eps' \Gamma_{p,f} = \cases{ 26 ~{\rm GeV} ~;~ 
(\nu_{\mu} ~{\rm and} ~\nu_e ~{\rm from} ~\pi^\pm ~{\rm and} ~\mu^\pm)
\cr 43 ~{\rm GeV} ~;~(\nu_{\mu} ~{\rm from} ~\mu^\pm)
\cr 60 ~{\rm GeV} ~;~ ({\gamma}) } \times \Gamma_{p,2.8}.
\label{observed-energy} \ea
Here $\Gamma_{p,2.8} = \Gamma_{p,f}/624$.

Note that additional pions may be produced via $pp\ra pn\pi^+$ and
$nn\ra np\pi^-$ inelastic interactions with similar cross-sections as
the inelastic $np$ interaction. The $nn$ process in particular may be
important for neutron destruction in the outflow \citep{rbr05}.
Neutrinos and photons produced from $nn$ and $pp$ processes, however,
are of much lower energy as the relative velocities between the
$n$-$n$ or $p$-$p$ components (e.g., arising from a variable outflow)
may not exceed that between $n$-$p$ in equation (\ref{n-p-bulk}).

\section{Neutrino emission}

The inelastic $n$-$p$ scattering opacity is $\tau'_{\pi} \simeq
\tau'_{np}\simeq 1$ at $r\sim R_{np} \sim R_{\tau}$ and each proton
produces, on average, a $\pi^\pm$ pair half of the time. The decay
neutrinos escape the plasma outflow as soon as they are created. The
neutrino and anti-neutrino flux rate from an SGRB at $z\simeq 0.1$
(corresponding to a luminosity distance $D_L \simeq 10^{27} D_{27}$ cm
for a standard cosmology with Hubble constant $H_o \simeq 70 ~{\rm
km~s^{-1} ~Mpc^{-1}}$) is
\ba {\dot N}_{\nu_{\mu}} \simeq {\dot N}_{{\bar \nu}_{\mu}} 
\simeq 2{\dot N}_{\nu_e} \simeq 2{\dot N}_{{\bar \nu}_e} \simeq 
\frac{\hat L}{4\pi D_L^2 \Gamma_{p,f} m_pc^2} 
\approx 5.7\times 10^{-6} \frac{{\hat L}_{49.8}}
{D_{27}^2 \Gamma_{p,2.8}} ~{\rm cm}^{-2}~{\rm s}^{-1}.
\label{nu-number} \ea

The neutrino-nucleon ($\nu N$) total cross-section for detection is
$\sigma_{\nu N} \approx 6\times 10^{-37}$ cm$^2$ at $\eps_{\nu}\approx
60$ GeV [see, e.g., Table 1. in \cite{gqrs98}]. Upcoming kilometer
scale water Cherenkov detectors such as {\it ANTARES} and {\it KM3NeT}
are designed to have sensitivity for detecting neutrinos of energy
$\gtrsim 50$ GeV. The total number of nucleons in a gigaton detector
is $N_{N} \approx 1.2\times 10^{39}$, which folded with the
cross-section and muon neutrino flux from a typical SGRB at $z\sim
0.1$ yields an event rate of
\ba {\dot N} (\nu_{\mu} + {\bar \nu}_{\mu} ~{\rm event}) 
\simeq 2 {\dot N}_{\nu_{\mu}} \sigma_{\nu N}N_{N} \approx 
0.01 ~{\hat L}_{49.8} D_{27}^{-2} \Gamma_{p,2.8}^{-1} 
~{\rm s}^{-1} {\rm Gton}^{-1} \label{nu-event-rate} \ea
per burst. Hence it is unlikely that upcoming neutrino telescopes
would be able to detect these neutrinos, based on our model as
dicussed here, from individual SGRBs. The number of events from
diffuse flux can be calculated assuming a burst rate of $\sim
300/4\pi$ yr$^{-1}$ sr$^{-1}$ (approximately $30\%$ of the long GRB
rate) distributed over the whole sky and an optimistic $2\pi$ sr
angular sensitivity of the detector as $(300/4\pi)\times 2\pi \times
{\dot N} (\nu_{\mu} + {\bar \nu}_{\mu} ~{\rm event}) t \approx 1$
yr$^{-1}$. Here, we have assumed $t\sim 1$ s as the typical duration
of the SGRBs.

\section{Photon emission}

The $\gamma$-ray emission from an SGRB, however, is more promising.
Although the plasma outflow becomes transparent to Compton scattering
at a radius $r\sim R_{\tau}\sim R_{np} \sim 10^8$ cm, not all the high
energy photons due to $\pi^0$ decay can escape. Only those created
below a skin depth of $R_{\tau}$ escape with a probability
\citep{bm00}
\ba P_{\pi^0} \simeq \tau'_{np} (R_{\tau}) \simeq 
R_{np}/R_{\tau} \approx 0.4. \label{pi0-fraction} \ea
The corresponding number flux of 60 GeV $\pi^0$-decay photons at Earth
is
\ba {\dot N}_{\gamma, \pi^0} \simeq 
\frac{P_{\pi^0} {\hat L}}{4\pi D_L^2 \Gamma_{p,f} m_pc^2} 
\approx 2.0\times 10^{-6} \frac{{\hat L}_{49.8}}
{D_{27}^2 \Gamma_{p,2.8}} ~{\rm cm}^{-2} ~{\rm s}^{-1}, 
\label{pi0-g-number}\ea
which is below the detection threshold of {\it GLAST} ``Large Area
Telescope'' (LAT), whose effective area is $\sim 10^4$ cm$^2$ in this
energy range \citep{gm99}. However, ground-based Cherenkov telescopes
with $\lesssim 100$ GeV threshold may detect these photons. In
particular, {\it Milagro} has an effective area of $\sim 5\times 10^5$
cm$^2$ for a zenith angle $\lesssim 15^\circ$ in this energy range
\citep{d04}.  For a yearly SGRB rate of $\sim$ 300/yr, this implies a
possible detection rate of $\lesssim$ 5 yr$^{-1}$, assuming a 90\%
duty cycle.

The high energy photons from $\pi^0$-decay created below $R_{\tau}$
interact with thermal photons of energy $\eps'_{\gamma,t} \simeq T'_o
(R_o/R_{\tau}) \approx 3.6$ keV, which are carried along with the
plasma.  The number density of these photons is high, and all
$\pi^0$-decay photons of energy $\eps'_{\gamma,i} \approx 100$ MeV
originating below the $\tau'_{\rm Th} \sim 1$ skin depth produce
$e^\pm$ pairs satisfying the condition $\eps'_{\gamma,t}
\eps'_{\gamma, i} \gtrsim m_e^2c^4$.  Assuming equal energy
distribution among final particles, charged pion decay $e^\pm$ and
$\gamma\gamma \ra e^\pm$ have a characteristic energy $\sim
35\gamma'_{\rm c.m.}$ MeV, and initiate electromagnetic cascades by
inverse-Compton (IC) scattering of thermal photons, creating new
generations of pairs which interact again with the thermal
photons. Contrary to the optically thick case of $\tau'_{\rm Th} \gg
1$, where all upscattered photons are absorbed by thermal photons
producing $e^\pm$ pairs, these cascades are not saturated
\citep{dkk99}. The upscattered photons eventually escape from the
photosphere with characteristic energy $\eps'_{\gamma,c} \approx
m_ec^2$ below the $e^\pm$ pair production threshold energy.

The fraction of the $\pi^\pm$ and $\pi^0$ energy carried by the
escaping $\gamma$-rays depends on the details. An order of magnitude
estimate for this fraction is $\xi_{\gamma} = 0.1\xi_{\gamma,-1}$
\citep{rbr05}.  Thus, the number flux of the escaping $\gamma$-rays of
energy
\ba \eps_{\gamma,c} \simeq \Gamma_{p,f} m_ec^2 
\approx 320 ~\Gamma_{p,2.8} ~{\rm MeV} \label{eps-mamma-c} \ea 
at Earth is
\ba {\dot N}_{\gamma,c} \simeq \frac{4 {\hat L}}
{4\pi D_L^2 \Gamma_{p,f} m_pc^2} \left( \xi_{\gamma} 
\frac{\gamma'_{\rm c.m.} m_{\pi}c^2} {m_ec^2} \right) 
\approx 8.6\times 10^{-4} \frac{{\hat L}_{49.8} \xi_{\gamma,-1}}
{D_{27}^2 \Gamma_{p,2.8}} ~{\rm cm}^{-2} ~{\rm s}^{-1}.
\label{ic-g-number} \ea 
The escaping Comptonized photon spectrum of the cascade is
$dN_{\gamma}/d\eps_{\gamma} \propto \eps_{\gamma}^{-q}$, where
$q\approx 5/3$ for $\eps_{\gamma} < \eps_{\gamma,c}$ and $q\approx
2.2$ for $\eps_{\gamma,c} < \eps_{\gamma} < \eps_{\gamma,\rm max}$
\citep{bd01}. The maximum photon energy is
$\eps_{\gamma,\rm max} \simeq \gamma_e^{'2} \eps'_{\gamma,t}
\Gamma_{p,f} \approx 20$ GeV. The fluence threshold for {\it GLAST} is 
$\sim 4\times 10^{-8}$ ergs cm$^{-2}$ for a short integration time
\citep{gm99}. Thus the Comptonized photon fluence 
$\eps_{\gamma,c}F_{\eps_{\gamma,c}} \simeq \eps_{\gamma,c}{\dot
N}_{\gamma,c}t \approx 4.4\times 10^{-7}$ ergs cm$^{-2}$ for a typical
SGRB of $t \sim 1$ s duration at $z\sim 0.1$ should be detectable by
{\it GLAST} (see Figure 2).

Apart from the $\pi^0$-decay photons of energy 60 GeV and $e^\pm$ pair
cascade photons of energy 320 MeV, the epectrum of radiation from the
photosphere of an SGRB would also contain thermal photons of peak
energy $\eps_{\gamma,b}\approx 2.82\eps'_{\gamma,t} \Gamma_{p,f}
\approx 6.4 \Gamma_{p,2.8}$ MeV (see Figure 2). The corresponding 
number flux on Earth of these photospheric photons of peak energy is
\ba {\dot N}_{\gamma,b} \simeq \frac{\hat L}{4\pi D_L^2 
\eps_{\gamma,b}} \approx 0.5 ~\frac{{\hat L}_{49.8}}
{D_{27}^2 \Gamma_{p,2.8}} ~{\rm cm}^{-2} ~{\rm s}^{-1}. 
\label{thermal-g-number} \ea

The photon fluxes at various energies discussed above are associated
with the jet photosphere, and last as long as the usual burst prompt
$\gamma$-ray emission. Hence, these emissions would be contemporaneous
with the emission at $\sim$ MeV associated with internal shocks or
other dissipation prcesses above the photopshere, aside from a small
time lag between the onset of the two events of the order of a
millisecond. We discuss emission from internal shocks next.

\subsection{Comparison with internal shock emission}

The internal shock model of plasma shells colliding at some radius
above the photosphere is the most commonly discussed mechanism to
produce the observed $\gamma$-rays in the fireball shock model. The
average bulk Lorentz factor of the colliding shells is $\Gamma_i
\simeq \Gamma_{p,f} \approx 624\Gamma_{i,2.8}$. With a millisecond
variability time $t_v =10^{-3}t_{v,-3}$ s, the internal shocks take
place at a radius $r_i \simeq 2\Gamma_i^2 ct_v \approx 2.3\times
10^{13}$ cm. Note that this is much smaller than the $\beta$-decay
radius $\sim 3.4\times 10^{15}$ cm. Hence, most neutrons have not had
time to decay until well beyond the internal shock radius.

The equipartition magnetic field in the shock region is $B= (2
\vareps_B {\hat L}/r_i^2 c)^{1/2} \approx 9.1\times 10^5$ G with an 
equipartition fraction $\vareps_B =0.1 \vareps_{B,-1}$. The
shock-accelerated electrons are assumed to follow a power-law energy
distribution $dN'_e/d\gamma'_e \propto \gamma_e^{'-\alpha}$ for
$\gamma'_e \ge \gamma'_{e,\rm min} \simeq \vareps_e m_p/m_e$ with
$\alpha \approx 2.4$, which results in synchrotron radiation with a
peak energy
\ba \eps_{\gamma,m} = \hbar c \Gamma_i 
\frac{3\gamma_{e,\rm min}^{'2}eB}{2m_e c^2} 
\approx 330 \left( \frac{\vareps_{e,-1}^2 \vareps_{B,-1}
{\hat L}_{49.8}}{\Gamma_{i,2.8}^2 t_{v,-3}^2} \right)^{1/2} 
~{\rm keV}. \label{pk-syn-energy} \ea
Here $\vareps_e = 0.1 \vareps_{e,-1}$ is the fraction of fireball's
kinetic energy transferred to the electrons by shocks. The
corresponding number flux at Earth of these peak synchrotron photons
is
\ba {\dot N}_{\gamma,m} \simeq \frac{{\hat L}\vareps_e}{4\pi D_L^2 
\eps_{\gamma,m}} \approx 1.0 \left( \frac{{\hat L}_{49.8} 
\Gamma_{i,2.8}^2 t_{v,-3}^2} {D_{27}^4 \vareps_{B,-1}} \right)^{1/2} 
~{\rm cm}^{-2} ~{\rm s}^{-1}. \label{syn-number} \ea
The typical synchrotron radiation spectrum from power-law distributed
electron energy is $dN_{\gamma}/d\eps_{\gamma} \propto
\eps_{\gamma}^{-(\alpha+2)/2} \sim \eps_{\gamma}^{-2.2}$ for
$\eps_{\gamma} > \eps_{\gamma,m}$ (see Figure 2). The observed typical
GRB spectra $dN_{\gamma}/d\eps_{\gamma} \propto \eps_{\gamma}^{-1}$,
is slightly different from the theoretical synchrotron spectrum with
an index $-3/2$.

We have plotted the energy spectra of synchroton radiation from the
internal shocks, and thermal and IC cascades from the jet photosphere
in Figure 2. We used $L=10^{50}L_{50}$ erg/s, $R_o=10^6$ cm and $\eta
= 316$ for all curves and $\xi_o=$ 1, 5 and 10 for the dashed, dotted
and dotted-dash curves respectively. The peak energy of thermal
photons ($\eps_{\gamma,b}$) decreases from 6.4 keV for $\xi_o=10$ to 3
keV for $\xi_o=1$ as the photosphere radius ($R_{\tau}$) in equation
(\ref{transparent-radius}) increases with decreasing $\xi_o$. The
variation of the peak energy ($\eps_{\gamma,c}$) of the photospheric
cascade is directly proportional to the variation of $\Gamma_{p,f}$
with $\xi_o$ and increases with $\xi_o$. We have normalized these
spectra by integrating the corresponding differential number fluxes
from $\eps_{\gamma,c}$ up to the respective maximum energy
($\eps_{\gamma, \rm max}$) and equating to the respective total number
flux in equation (\ref{ic-g-number}) with fixed $\xi_{\gamma} =0.1$,
the fraction of the total energy carried by the pions. The variation
of the synchrotron peak energy ($\eps_{\gamma,m}$), in the range
330-400 keV, with $\xi_o$ can be explained similarly. We have
normalized the synchrotron spectra by integrating the corresponding
differential number fluxes from $\eps_{\gamma,m}$ up to the energy
$10\eps_{\gamma, m} \sim$ few MeV, the typical energy range for the
{\it BATSE} and {\it BAT} detectors, and equating to the respective
total number flux in equation (\ref{syn-number}). We have plotted an
extension of the synchrotron spectra up to 1 GeV, in order to compare
them with other spectral components.

The synchrotron spectrum does not extend very far above
$\eps_{\gamma,m}$, for the parameters used here, from a calculation of
the maximum electron Lorentz factor. The energy at which the
Rayleigh-Jeans portion of the thermal photospheric photons start to
dominate over the synchrotron spectrum is $\eps_{\gamma}
\approx 2$ MeV for $\xi_o=5$ and 10 curves in Figure 1. 
(For $\xi_o=1$, the thermal component is small over the whole energy
range.) Above this, besides the Comptonized photospheric cascade
spectrum, there could also be an IC scattered component of the
synchrotron photons (SSC), whose importance is characterized by the
parameter $Y= [-1+ (1+ 4\vareps_{e}/\vareps_B)^{1/2}]/2 \approx 0.6$
for our present model.  The corresponding minimum IC peak is
$\gamma_{e, \rm min}^{'2}\sim 4\times 10^4$ higher than the
synchrotron peak, at a photon energy $\eps_{\gamma,m, \rm IC}\sim 10$
GeV. Hence an SSC component is not important in this case
($Y<1$). Thus, at energies above which the sycnhrotron spectrum cuts
off, the Comptonized photospheric cascade photon spectrum would be
clearly distinguishable by {\it GLAST} (see Figure 1). Although for
some choice of parameters an SSC contribution could become significant
at energies $\gtrsim 10$ GeV, this is expected to be at energies much
higher than the peak of the Comptonized cascade peak
$\eps_{\gamma,c}\sim 300$ MeV. The maximum SSC photon energy, in the
Thomson limit $\gamma'_e \ll m_ec^2/\eps_{\gamma, m}$, would be
$\eps_{\gamma,\rm IC} \ll m_e^2 c^4 \Gamma_i^2/ \eps_{\gamma,m}
\approx 300$ GeV. At energies well above $\sim 10$ GeV the two
components (Comptonized photospheric cascade and SSC) may be hard to
distinguish, but in the range of $\sim$5 MeV - 10 GeV the Comptonized
photospheric cascade should dominate. Below the peak $\eps_{\gamma}
\lesssim \eps_{\gamma,c}$, the Comptonized photospheric cascade
component is lower than the thermal photospheric peak, and at even
lower energies it is also lower than the synchrotron component.

\section{Discussion}

If SGRBs are produced as a result of double neutron star or neutron
star-black hole binary mergers, the neutron-rich outflow may result
naturally in a high proton Lorentz factor $\Gamma_{p,f}\gtrsim 600$.
Besides the usual non-thermal (e.g. internal shock) synchrotron
spectrum, this implies the presence of a thermal photospheric
component peaking around $\sim 10$ MeV. The $n$-$p$ decoupling will
lead to higher energy photons ($\gtrsim 0.1$ GeV) from neutral and
charged pion decay cascades, and neutrinos ($\gtrsim 30$ GeV) directly
from pion and muon decays. There may be additional X-ray emission when
the neutron-decay proton shell with initial lower Lorentz factor
impact with the decelerating proton shell initially moving with a high
Lorentz factor \citep{da06}. The afterglow light curves can also be
affected by the neutron decay \citep{dkk99b,rbr05}. Here, we have
concentrated on the prompt ($t\lesssim 2$ s) high energy (MeV to tens
of GeV) signatures of neutron-rich outflows. For short bursts at
$z\sim 0.1$ the $\sim 50$ GeV neutrinos are unlikely to be detectable
with currently planned Gigaton Cherenkov detectors. However, the pion
decay photons at $\sim 60$ GeV escaping from whithin a $\tau'_{\rm
Th}\sim 1$ skin-depth of the photosphere should be detectable with
high area, high duty cycle Cherenkov detectors such as {\it Milagro},
at an estimated rate of 5 bursts per year. The pion decay photons
created below the photosphere will result in a Comptonized cascade
spectrum which, peaking in the range $\gtrsim 0.3$ GeV, dominates the
non-thermal shock synchrotron-IC spectrum, and should be detectable
with {\it GLAST}.  Both the prompt GeV photon components discussed
here are essentially contemporaneous with the usual MeV range short
GRB prompt emission, $t_\gamma\lesssim 2$ s.  Their detection or lack
thereof could constrain the neutron content of the outflow and the
nature of the short burst progenitors.

\acknowledgements{We thank the referee for comments which have 
helped us to improve the paper. Research supported in part through NSF
AST 0307376 and NASA NAG5-13286.}

\clearpage

\begin{figure} 
\plotone{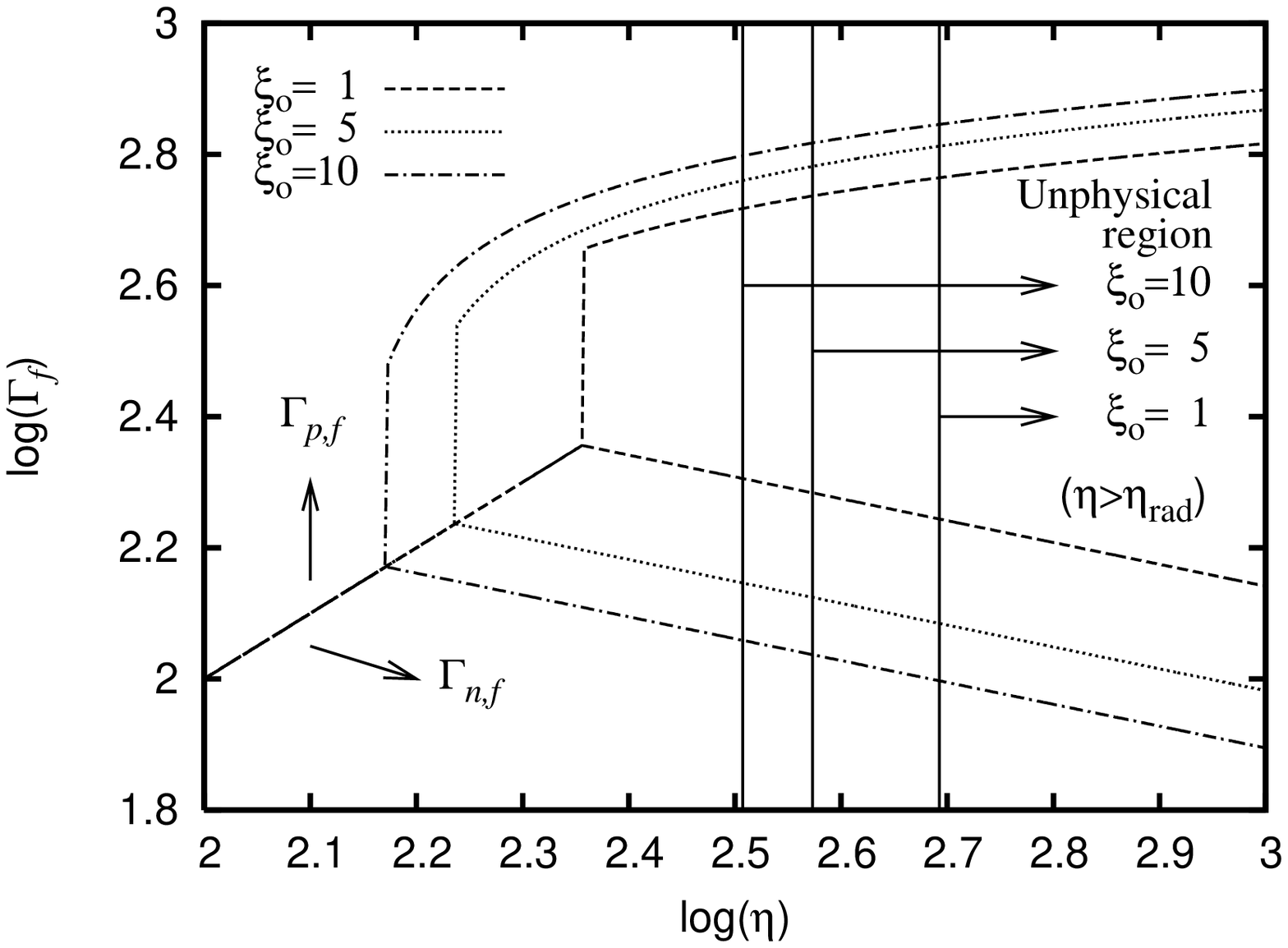}
\caption{The final bulk Lorentz factor of the proton ($\Gamma_{p,f}$) 
and neutron ($\Gamma_{n,f}$) components, using equations
(\ref{final-neutron-bulk}) and (\ref{final-proton-bulk}) respectively,
in the GRB jet outflow as functions of the baryon-loading parameter
($\eta$). The dashed, dotted and dotted-dash curves corresponds to the
initial neutron to proton ratio $\xi_o = n'_n/n'_p =$ 1, 5 and 10
respectively. The other parameters used are $L=10^{50}L_{50}$ erg/s
and $R_o=10^6$ cm for all curves. For $\eta < \eta_{np}$, defined in
equation (\ref{np-entropy}), $\Gamma_{p,f} =\Gamma_{n,f} = \eta$ as
shown by the single line. The protons and neutrons are held together
in the outflow with leptons and photons in this case. For $\eta >
\eta_{np}$ the protons and neutrons decouple as shown by the splitting
of the curves at $\eta = \eta_{np}=$ 148, 172 and 227, and
$\Gamma_{p,f} > \Gamma_{n,f}$. Note that the $n$-$p$ decoupling takes
place for lower $\eta$ as $\xi_o$ increases with other parameters
fixed. The vertical lines at $\eta=$ 322, 374 and 493 correspond to
the limiting value of $\eta=\eta_{\rm rad}$ defined in equation
(\ref{rad-entropy-def}).}
\end{figure}

\clearpage

\begin{figure} 
\plotone{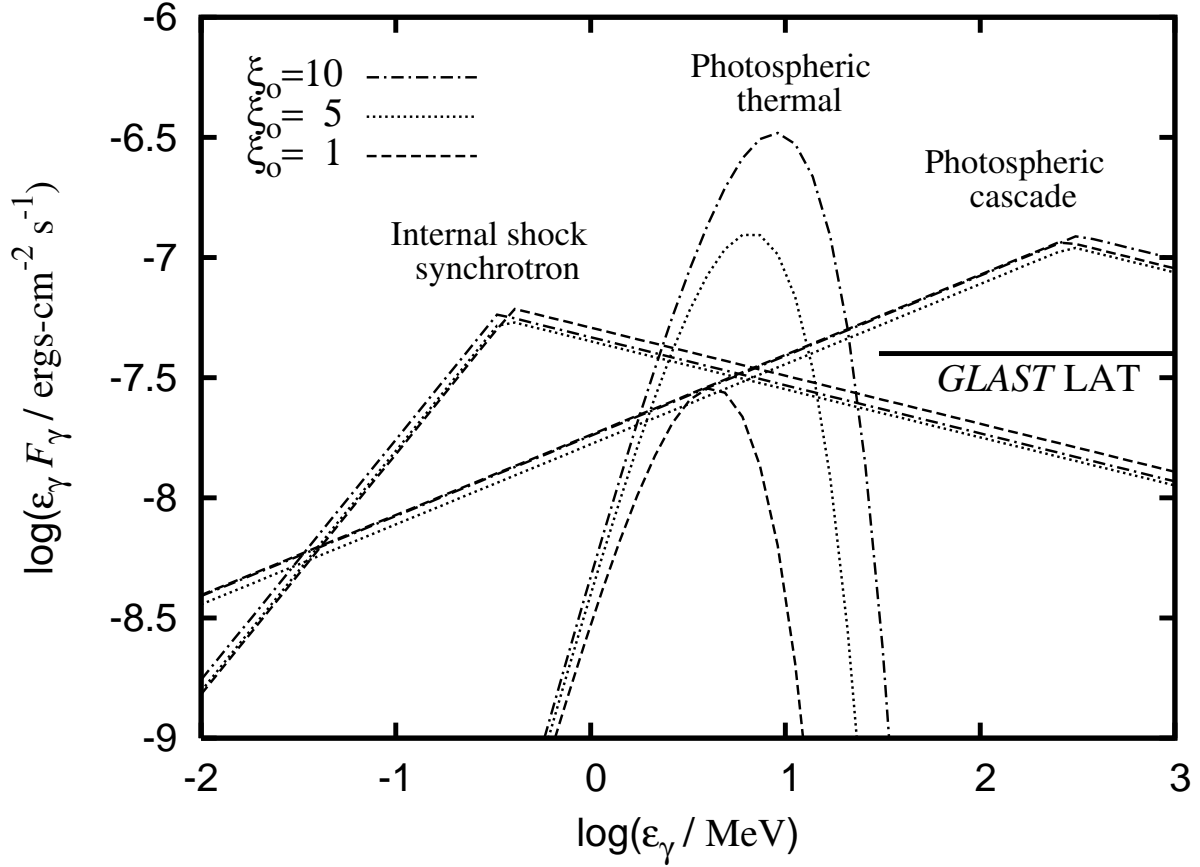}
\caption{Photon spectra of the expected photospheric signals (thermal 
and IC cascade by pion decay electrons and photons) and the
synchrotron spectrum from internal shocks, compared to the {\it GLAST}
LAT sensitivity. The dashed, dotted and dotted-dash curves corresponds
to the initial neutron to proton ratio $\xi_o = n'_n/n'_p =$ 1, 5 and
10 respectively as in Figure 1. The other parameters used are
$L=10^{50}L_{50}$ erg/s, $R_o=10^6$ cm and $\eta = 316$ for all
curves. To calculate the internal shock synchrotron spectra we used a
fixed variability time $t_v=10^{-3}$ s. The peaks of the spectra are
correlated with the final bulk Lorentz factor of the proton outflow:
$\Gamma_{p,f}=$ 519, 572 and 624 respectively for $\xi_o=$ 1, 5 and 10
(see main text for more details).}
\end{figure}

\end{document}